\documentclass[reqno]{amsart}
\def\be{\begin{equation}}
\def\ee{\end{equation}}
\def\ba{\begin{array}}
\def\ea{\end{array}}
\def\bea{\begin{eqnarray}}
\def\eea{\end{eqnarray}}

\def\noi{\noindent}

\begin{document}

\vspace{-4truecm} %
{}\hfill{DSF$-$05/2007}%

\vspace{1truecm}

\title{Majorana and the quasi-stationary states in
Nuclear Physics}

\author{E. Di Grezia}%
\author{S. Esposito}
\address{{\it E. Di Grezia}: Dipartimento di Scienze Fisiche,
Universit\`a di Napoli ``Federico II'' \& I.N.F.N. Sezione di
Napoli, Complesso Universitario di M. S. Angelo, Via Cinthia,
80126 Napoli ({\rm digrezia@na.infn.it})}%
\address{{\it S. Esposito}: Dipartimento di Scienze Fisiche,
Universit\`a di Napoli ``Federico II'' \& I.N.F.N. Sezione di
Napoli, Complesso Universitario di M. S. Angelo, Via Cinthia,
80126 Napoli ({\rm Salvatore.Esposito@na.infn.it})}%

\begin{abstract}
A complete theoretical model describing artificial disintegration
of nuclei by bombardment with $\alpha$-particles, developed by
Majorana as early as in 1930, is discussed in detail alongside the
basic experimental evidences that motivated it. By following the
quantum dynamics of a state resulting from the superposition of a
discrete state with a continuum one, whose interaction is
described by a given potential term, Majorana obtained (among the
other predictions) the explicit expression for the integrated
cross section of the nuclear process, which is the direct
measurable quantity of interest in the experiments. Though this is
the first application of the concept of quasi-stationary states to
a Nuclear Physics problem, it seems also that the unpublished
Majorana's work anticipates by several years the related seminal
paper by Fano on Atomic Physics.
\end{abstract}

\maketitle


\section{Nuclear Physics studies in Rome}

\noi After the advent of Enrico Fermi to the chair of Theoretical
Physics at the University of Rome in 1926, a group of young people
was started to form around him, mainly due to the decisive support
of Orso Mario Corbino, the Director of the Institute of Physics in
Rome. Aimed to a powerful relaunch of Physics in Italy, upon
Fermi's suggestion Corbino firstly hired Franco Rasetti as his
assistant, which was an excellent experimentalist who studied with
Fermi in Pisa and later worked with him in Florence. Then, after a
famous appeal by Corbino to the students of Engineering, in order
to convince the most brilliant ones to register in Physics, Emilio
Segr\`e and Edoardo Amaldi decided to join the rising group of
Fermi, though they were still students. After some months, at the
end of 1927, Ettore Majorana passed as well to Physics, the
details of the first meeting by Majorana and Fermi being well
known \cite{[Segre]}\cite{[Amaldi]}\cite{[DiGrezia]}.

At that time, Fermi was working on the statistical model of atoms
(the Thomas-Fermi model), and the activities of his and Rasetti's
group were thus devoted to researches on Atomic Physics. However,
after having obtained several significant results in atomic and
molecular spectroscopy, around 1930 some people in the group
realized that such a field could offer no longer any great
prospect, and Fermi himself predicted that the interest would have
shifted from the study of the external parts of the atom to its
nucleus. No general consensus, however, created initially inside
the group. Segr\`e, for example, recalls that \cite{[Segre]}:
\begin{quote}
My personal reaction was that we had just learned spectroscopic
techniques, with which we were reaping good results, and that we
might persist in that field a little longer. I was, however, open
to Fermi's arguments. Amaldi and Rasetti also had their points of
view, and we had long, lively discussions on the subject. As was
to be expected, Fermi's ideas prevailed, although everybody was
left free to do what he liked best. Thus I continued to work
experimentally on spectroscopy until we started on neutron work.
However, we all increased our reading on nuclear subjects.
\end{quote}
The spectroscopic activities, in fact, continued side by side in
the acquisition of theoretical knowledge and experimental
technologies required by Nuclear Physics, till 1934, when the
Fermi's group discovered the radioactivity induced by neutrons and
the important properties of slow neutrons \cite{[DeGregorio]}. As
a bridge between spectroscopy and Nuclear Physics, we find two
theoretical works of 1930 \cite{[momentimagnetici]} and 1933
\cite{[struttureiperfini]}. In the first of these, performed by
Fermi, the magnetic moments of sodium, rubidium and cesium nuclei
were deduced from the hyperfine atomic structures observed in the
spectra of these elements. Instead, in the second one, Fermi and
Segr\`e performed quite exhaustive study on the nuclear magnetic
moments of 17 element, aimed to show that hyperfine structures of
atomic spectra could be completely explained by the nuclear
magnetic moment, and that there were no other nuclear forces at
play. The work by Fermi and Segr\`e resulted into some involved
calculations, who required the computation of the coupling
constants of single electrons and that of the perturbations of the
electronic terms, and the authors themselves explicitly
acknowledged some help on these by Ettore Majorana. According to
direct testimonies \cite{[Segre]}\cite{[Amaldi]} and, mainly, to
what reported in his research notebooks (see, for example,
\cite{[Volumetti]}), Majorana actively participated in scientific
discussions inside the Fermi's group, but we are left with only
one published paper of him on Nuclear Physics topics
\cite{[Kern]}, dealing with the known Majorana-Heisenberg exchange
forces in nuclei. Practically nothing is known, until now, of his
previous works on these matters; here we will focus just on this.
We firstly point out that, indeed, Majorana was the {\em first} in
Rome to study nuclear physics, at least since 1929 when (on July
6) he defended his master thesis on \"The quantum theory of
radioactive nuclei". However, quite unexpected, he continued to
study such topics for several years, also independently from the
main researches of the Fermi's group, till his famous theory of
nuclear exchange forces of 1933. In the following we will report
on relevant studies conducted mainly on some unpublished notes by
Majorana ranging on this time period, revealing previously unknown
interesting results obtained by him on Nuclear Physics topics. The
resulting picture, properly included in its contemporary
historical framework, points out once more the extraordinary
abilities of the Italian physicist.

\section{Exploring the nucleus' secrets with $\alpha$ particles}

\noi The first steps of Nuclear Physics were moved in the realm of
radioactive phenomena, implying the emission of some "rays" from
several substances like uranium. It was Rutherford to realize that
such rays were made of two different components, the first one,
easily absorbed by matter, being denoted as $\alpha$ radiation,
while the second one, much more penetrating, as $\beta$
radiation,. Radioactive phenomena involving $\alpha$ emission were
readily recognized to induce a profound modification of the atom
(or, as was realized later, of the nucleus), due to the large mass
of the emitted $\alpha$ particles. The extremely high values of
the energies released in these processes, furthermore, led
Rutherford himself in 1904 to postulate that the causes of
radioactivity should be searched inside the atom and later, in
1911, to interpret the famous experiment of Geiger and Marsden, on
the $\alpha$ bombardment of atoms, in terms of his nuclear atomic
model. Contextually, artificial radioactivity induced by particle
irradiation was discovered and, in particular, the emergence of
protons as disintegration products of nitrogen bombarded with
$\alpha$ particles, led to the conclusion that the masses of
nuclei were primary due also to protons, besides $\alpha$
particles. The problem of the possible presence of electros inside
the nucleus, driven by the fact that such light particles were
emitted in $\beta$-decays, remained instead controversial and
unsolved until the discovery of the neutron by Chadwick in 1932
and the formulation of the Fermi theory of $\beta$-decay one year
later (a detailed account of this question may be found in
\cite{[elettroninucleari]}). Here we only quote, in passing, a
far-seeing intuition of Majorana on this problem in 1929:
\begin{quote}
It seems to me that the problem of the aggregation of protons and
electrons in nuclei cannot have solutions, though approximate,
until the problem of the constitution of protons and electrons
themselves has been solved. This is for a very simple reason: the
size of complex nuclei... are of the same order of that of
electrons (obviously, this is evaluated on classical grounds.
Quantum mechanics has not brought, and {\em cannot brought by
itself}, any light on this question, since no relation between $e,
h, m$ can hold due to dimensional reasons). Although such
statements are largely vague, we can expect that something is
hidden in them. \cite{[GentileMa;29]}
\end{quote}
Much information on nuclear structure and dynamics came out, in
1920's and 1930's, from experiments on nuclei bombarded with
$\alpha$ particles (as a later example, we mention the discovery
of the neutron anticipated by researches on $\alpha$ irradiation
on boron and berillium), as one could obviously expect due to the
peculiar transformations induced by such massive particles. On the
theoretical side, the first important achievement was due to
Gamow, that provided the theory for the spontaneous $\alpha$
emission from a nucleus in the framework of Quantum Mechanics
\cite{[GamowAdecay]}. Just after the appearance of the Gamow
paper, in 1928 Majorana, who was studying Nuclear Physic (as well
as some Atomic Physics topics at the same time) in preparing his
master thesis supervised by Fermi, considered also the inverse
process of absorption of an $\alpha$ particle by a radioactive
nucleus \cite{[Volumetti2.34]}. It is quite instructive to see how
Majorana approached such a problem:
\begin{quote}
Let us consider the emission of an $\alpha$ particle by a
radioactive nucleus and assume that such a particle is described
by a quasi-stationary wave. As Gamow has shown, after some time
this wave scatters at infinity. In other words, the particle
spends some time near the nucleus but eventually ends up far from
it. We now begin to study the features of such a quasi-stationary
wave, and then address the inverse of the problem studied by
Gamow. Namely, we went to determine the probability that an
$\alpha$ particle, colliding with a nucleus that has just
undergone on $\alpha$ radioactive transmutation, will be captured
by the nucleus of the element preceding the original one in the
radioactive genealogy.
\end{quote}
After lengthy calculations (the complete paper may be found, in
English, in \cite{[Volumetti2.34]}), Majorana indeed succeeded in
obtaining the expression for the absorption probability (in two
different ways: by using mechanical or thermodynamical arguments),
\begin{equation}
\frac{n}{N}=\frac{2\pi^2\hbar^3}{m^2vT}
\end{equation}
(T being the life-time and $m, v$ the mass and velocity of the
$\alpha$ particle, respectively), by relating it to nuclear
$\alpha$-decay quantities and showing that such probability is
``completely independent of any hypothesis on the form of the
potential near the nucleus, and that it only depends on the
lifetime T." Intriguing results were then obtained in 1929 from
the artificial disintegrations of nuclei by means of
$\alpha$-particles. It was already known that, when bombarding
certain elements by $\alpha$-particles, protons were emitted as a
result of the disintegration of the considered nuclei. On the
basis of Blackett's photographs \cite{[Pollard]} of the
disintegration of the nitrogen nucleus, it was assumed that in
such a process the $\alpha$-particle is absorbed by the atomic
nucleus, which is thereby transformed into a nucleus of the
element of the next higher atomic number. According to this view,
by applying energy-momentum conservation laws, simple relations
between the energy of the incident $\alpha$ particle and that of
the emitted proton should hold and, furthermore, the change of
energy from the original to the final nucleus should be a fixed
amount. In fact, if we denote with $-E^0_p$ an energy level of a
proton in a given nucleus and with $-E^0_\alpha$ that for an
$\alpha$-particle, when such a particle with energy $E_\alpha$ is
captured by the nucleus, the energy of the proton emitted in the
disintegration will be $E_p=E_\alpha+ E^0_\alpha -E^0_p$
(neglecting the small kinetic energy of the residual nucleus).
Thus, in general, protons with definite energy will be emitted
during this process, so that their energy spectrum is a discrete
one. However, Rutherford and Chadwick in 1929 \cite{[RuthChad]}
clearly showed that, at least in some cases (for example in the
transition between aluminium and silicon), this is not true, and
the change of energy was not always the same. This should lead, as
a consequence, to assume that the energy levels of the protons in
the $\alpha$ particles in the nucleus are not well defined, in
contrast to what emerged from the application of Quantum Mechanics
to the nuclear structure. Such evidences were later confirmed by
Chadwick, Constable and Pollard \cite{[Pollard]} but, in the
meantime, a novel idea for interpreting them came out. Chadwick
and Gamow \cite{[ChadGamow]}, in fact assumed that, in some cases,
the disintegration of the nucleus may occur by the ejection of a
proton {\em without} capture of the $\alpha$-particle. In this
occurrence, the energy of the incident $\alpha$-particle is
distributed between the emitted proton and the escaping
$\alpha$-particle (neglecting the recoiling nucleus), so that the
disintegration protons may have any energy between $0$ and
$E_\alpha -E^0_p$, thus explaining the experimental evidences. As
a consequence, if both types of disintegration (with and without
$\alpha$-capture) take place, in the simplest case the energy
spectrum of the emitted protons is composed of a continuous
spectrum and, after its endpoint, a line. The importance of the
identification of the continuous spectrum was readily recognized
by Chadwick, Constable and Pollard themselves: ``It is most
important to find the continuous spectrum of protons corresponding
to disintegration without capture, for this gives immediately the
level of the proton in the nucleus." \cite{[Pollard]}. The
theoretical interpretation of the results from $(\alpha ,p)$
reactions was mainly due to Chadwick and Gamow but, to the best of
our knowledge, no dynamical theory for those processes describing
the superposition of a continuous spectrum and a discrete level
was ever published. Recently, however, we have realized that such
a theory was effectively elaborated by Majorana in 1930, although
he never reported it in any journal, being contained in his
personal notebooks. In the following we will give a detailed
account of the Majorana theory (see \cite{[Volumetti4.28]}),
pointing out the interesting, unknown results obtained by the
Italian physicist.

\section{Majorana general theory for quasi-stationary states}

\subsection{First problem}

\noi Let us consider, in general, a physical system for which a
discrete state $\psi_0$ exists with energy $E_0$ together with a
continuum one $\psi_W$ with energy $E_0 +W$ (the state $\psi_W$ is
normalized with respect to $dW$). The perturbation linking the
discrete state with the continuum ones is denoted by Majorana as:
\begin{equation}
I_W =\int \bar{\psi}_0 H_p \psi_W d\tau , \label{1}
\end{equation}
where $H_p$ is the perturbation potential and $d\tau$ is the
volume element. The problem is that of obtaining the perturbed
eigenfunctions $\psi'_W$ of the total hamiltonian $H$:
\begin{equation}
H\psi'_W =(E+W)\psi'_W, \label{2}
\end{equation}
with
\begin{eqnarray}
&& H\psi_0=E_0\psi_0+\int \bar{I}_W  \psi_W dW, \label{3} \\
&&H\psi_W=(E_0 + W)\psi_W +I_W \psi_0. \label{4}
\end{eqnarray}
Majorana finds the analytic solutions for the eigenfunctions,
which are expressed as follows:
\begin{equation}
\psi'_W= \frac{1}{N_W}\left\{\psi_0 -\int \bar{I}_{W'}
\frac{\psi_{W'}}{W'-W}dW' + a \psi_W \right\} \label{5}
\end{equation}
where $N_W =\sqrt{|a|^2 +|b|^2}$ is a normalization factor written
in terms of the parameters:
\begin{eqnarray}
&& a=I_W^{-1}\left\{W + \int |I_W|^2  \frac{dW'}{W'-W}\right\}, \label{6} \\
&&b=\pi I_W \label{7}
\end{eqnarray}
(here and below, the integrals are assumed to be evaluated at
their principal value). \\ In such a way, the discrete state
$\psi_0$ may be expanded in terms of the perturbed eigenfunctions
$\psi'_{W}$ as:
\begin{equation}
\psi_0 =\int \frac{1}{N_W} \psi'_W dW. \label{8}
\end{equation}
In order to get explicit results, Majorana then considers the
approximate case where terms higher than the second in $I_W$ are
neglected, so that the quantities:
\begin{equation}
I_W\simeq I, \qquad \qquad \int |I_W|^2  \frac{dW'}{W'-W}\simeq k
\label{9}
\end{equation}
may be regarded as constants. By setting $W=\varepsilon -k$ we
have $a=\varepsilon/I$, $b=\pi I$, $N =\sqrt{\varepsilon^2/|I|^2 +
\pi^2 |I|^2}$ and the perturbed eigenfunctions are:
\begin{equation}
\psi'_W \simeq \frac{1}{\sqrt{\varepsilon^2/|I|^2 + \pi^2
|I|^2}}\left\{\psi_0 - \bar{I} \int  \frac{\psi_{W'}}{W'-W}dW' +
\frac{\varepsilon}{I}\psi_W\right\}. \label{10}
\end{equation}
Now, let us assume that at time $t=0$ the system is in the state
$\psi_0$. If the time-dependent factor of $\psi'_W$ is taken to
be:
\begin{equation}  \label{11}
e^{-iEt/\hbar}=e^{-i(E_0-k)t/\hbar}e^{-i\varepsilon t/\hbar},
\end{equation}
the time evolution of the state of the system is given by:
\begin{equation}  \label{12}
\psi=e^{-i(E_0-k)t/\hbar}\left\{e^{-t/2T}\psi_0 + \int
\frac{\bar{I}}{\varepsilon + i\pi |I|^2}\left(e^{-i\varepsilon
t/\hbar}- e^{-t/2T}\right)\psi_Wd\varepsilon\right\}.
\end{equation}
Here,
\begin{equation}  \label{13}
\frac{1}{T}=\frac{2\pi}{\hbar}|I|^2
\end{equation}
gives the transition probabilities per unit time between the
initial (discrete) state $\psi_0$ and the states $\psi_W$.

\subsection{Second problem}

\noi The next problem considered by Majorana was to study the
transitions between one discrete state $\psi_0$ with energy $E_0$
and two continuum ones $\psi_W$ and $\phi_W$, both with energy
$E_0+ W$. In a way analogous to the previous one, the two
perturbations are written as:
\begin{eqnarray}
&& I_W =\int \bar{\psi}_0 H_p \psi_W dW,  \label{14} \\
&&L_W =\int \bar{\psi}_0 H_p \phi_W dW,  \label{15}
\end{eqnarray}
and
\begin{eqnarray}
H\psi_0 &=& E_0\psi_0 + \int \bar{I}_W \psi_W dW + \int \bar{L}_W \phi_W dW, \label{16} \\
H\psi_W &=& (E_0 + W)\psi_W + I_W \psi_0 ,  \label{17} \\
H\phi_W &=& (E_0 + W)\phi_W + L_W \phi_0  \label{18} .
\end{eqnarray}
Now, for any value of $W$, we have two stationary states $Z_W^1$
and $Z_W^2$,
\begin{eqnarray}
&&HZ_W^1 = (E_0 + W)Z_W^1  \label{19} \\
&&HZ_W^2 =(E_0 + W)Z_W^2,  \label{20}
\end{eqnarray}
that can be chosen to be orthogonal between them and normalized in
such a way that:
\begin{eqnarray}
Z_W^1 &=& \frac{1}{N'_W}\left\{\psi_0 +
\varepsilon_W\frac{\bar{I}_W}{Q_W^2}\psi_W +
\varepsilon_W\frac{\bar{L}_W}{Q_W^2}\phi_W \right. \nonumber \\
& & \left. -  \int \frac{\bar{I}_{W'}\psi_{W'}}{W'-W}dW'- \int
\frac{\bar{L}_{W'}\phi_{W'}}{W'-W}dW'\right\}, \label{21}  \\
Z_W^2 &=& \frac{L_W\psi_W}{Q_W}-\frac{I_W\phi_W}{Q_W}. \label{22}
\end{eqnarray}
where:
\begin{eqnarray}
&&\varepsilon_W = W + \int |I_{W'}|^2  \frac{dW'}{W'-W} + \int
|L_{W'}|^2  \frac{dW'}{W'-W} =W+K_W  \label{23} \\
&& Q_W =\sqrt{|I_{W}|^2 + |L_{W'}|^2},  \label{24} \\
&& N'_W =\sqrt{\frac{\varepsilon^2_W}{Q_W^2}+ \pi Q_W^2}.
\label{25}
\end{eqnarray}
Since $Z_W^2$ are orthogonal to $\psi_0$, this can be expanded by
using only the states $Z_W^1$:
\begin{equation}
\psi_0=\int \frac{Z_W^1}{N_W'}dW.  \label{26}
\end{equation}
By using analogous approximations to that introduced above,
Majorana then finds the time evolution of the state of the system
which is initially in the state $\psi_0$:
\begin{eqnarray}
\psi &=& e^{-i(E_0-k)t/\hbar}e^{-t/2T}\psi_0 \nonumber
\\ &+& \bar{I} \int \frac{\psi_We^{-iEt/\hbar}}{W + K + i\pi(|I|^2
+ |L|^2) }\left(1 -e^{i(W+K)t/\hbar}e^{-t/2T}\right)dW+\nonumber \\
&+& \bar{L} \int \frac{\phi_W e^{-iEt/\hbar}}{W + K + i\pi(|I|^2 +
|L|^2) }\left(1 -e^{i(W+K)t/\hbar}e^{-t/2T}\right)dW, \label{27}
\end{eqnarray}
where
\begin{equation}
 \frac{1}{T}= \frac{2\pi}{\hbar}(|I|^2
+ |L|^2).  \label{28}
\end{equation}
The transition probability per unit time between $\psi_0$ and the
states $\psi_W$ or $\phi_W$ is $2\pi|I|^2/\hbar$ or
$2\pi|L|^2/\hbar$, respectively.

\subsection{Third problem}

\noi Now let us suppose that the system is initially in the
continuum state $\psi_W$ and consider the relative probability
that at time $t$ the system is in the state $\psi_0$, or in the
states $\psi_W$, or in the states $\psi'_W$ with $W'\neq W$. First
of all, we note that Majorana ``prefers" to use the concept of
``number of systems" in a given state rather than that of the
relative probability for the system to be in that state. Although
the initial state $\psi_W$ is not a stationary state and
represents an infinite number of systems, only a finite of them
has an energy that differs from $E_0 + W$ by a finite amount, so
that we can effectively expect only transitions to states next to
$\psi_W$ and $\phi_W$. In such framework, the initial state may be
expanded in terms of the stationary states $Z_W^1$ and $Z_W^2$,
and its time evolution at time $t$ is written (by using the usual
above approximation) as:
\begin{eqnarray}
\psi & = & \frac{1}{N'_W} \frac{\varepsilon I}{|I|^2 +
|L|^2}e^{-iEt/\hbar}Z_W^1+ I \int e^{-iE't/\hbar}\frac{Z_{W'}^1 }{N'_{W'}(W'-W)}\nonumber \\
&+& \frac{\bar{L}} {|I|^2 + |L|^2}e^{-iEt/\hbar}Z_W^2 \label{29}
\end{eqnarray}
$(E=E_0 +W,\,\,E' = E_0 + W')$. Here Majorana also notes that, in
order to regularize the integral in (\ref{29}), it is convenient
(above and in the following) to use the replacement:
\begin{equation}  \label{30}
\frac{1} {W'-W}\rightarrow \frac{W'-W} {(W'-W)^2 + \alpha^2}
\end{equation}
and take then the limit $\alpha\rightarrow 0$ at the end of
calculations. The expression of $\psi$ in terms of the unperturbed
states $\psi_0, \psi_W, \phi_W$ may be obtained by using the
expressions of $Z_W^1, Z_W^2$ in Eqs. (\ref{21}) (\ref{22}).
However, Majorana points out that, for later times $t>0$, it is
convenient to express $\psi$ as a sum of two terms $\psi_1$ and
$\psi_2$, such that $\psi_1 + \psi_2 =\psi_W$ at $t=0$ and that
$\psi_1$ substantially describes the phenomenon for sufficiently
large times, while $\psi_2$ is one of the discrete states of the
form given in (\ref{27}). He then finds:
\begin{equation}  \label{31}
\psi = \psi_1 + \psi_2
\end{equation}
\begin{eqnarray}
\psi_1 &=& e^{-iEt/\hbar}\psi_W + \frac{I}{\varepsilon +i\pi Q^2}
\,e^{-iEt/\hbar}\psi_0 \nonumber \\
&-& \frac{I}{\varepsilon +i\pi Q^2}\int \frac{\bar{I}\psi_{W'}
+\bar{L}\phi_{W'}}{\varepsilon'-\varepsilon}e^{-iE't/\hbar}\left(1-e^{-i(E-E')t/\hbar}
\right)dE',
\nonumber \\
\psi_2 &=& - \frac{I}{\varepsilon +i\pi Q^2}
\left\{e^{-i(E_0-K)t/\hbar}e^{-t/2T}\psi_0\right.
\nonumber \\
&+& \left.\int \frac{\bar{I}\psi_{W'} +\bar{L}\phi_{W'}}
{\varepsilon'+i\pi
Q^2}e^{-iE't/\hbar}\left(1-e^{-i(E'-E_0+K)t/\hbar}e^{-t/2T}\right)dE'\right\}
\nonumber
\end{eqnarray}
($Q=\sqrt{|I|^2 + |L|^2}$, $\varepsilon =E-E_0 +K$, $\varepsilon'
=E'-E_0 +K $) with
\begin{equation} \frac{1}{T}=
\frac{2\pi}{\hbar}Q^2.  \label{32}
\end{equation}
For sufficiently large times, the number of transitions per unit
time from $\psi_W$ to the states $\psi_{W'}$ with energy $W'$
close to $W$ is dominated by the resonance term
$1/(\varepsilon'-\varepsilon)$ in the expression for $\psi_1$.
Thus, the number $A$ of transitions to states $\psi_{W'}$ or that
$B$ to states $\phi_{W'}$ is estimated as:
\begin{equation}  \label{33}
A=\frac{2\pi}{\hbar}|I|^2\frac{|I|^2}{\varepsilon^2 +\pi^2
Q^2},\,\,\,\;\;
B=\frac{2\pi}{\hbar}|L|^2\frac{|I|^2}{\varepsilon^2 +\pi^2 Q^2},
\end{equation}
respectively.

\subsection{Fourth problem}

\noi In order to describe several practical applications, Majorana
finally introduces some linear combinations of the states $\psi_W$
and $\phi_W$, having a definite physical meaning:
\begin{equation}  \label{34}
\psi_{W}= u_W^1+ u_W^2,\,\,\,\,\;\;\;\phi_{W}= v_W^1+ v_W^2,
\end{equation}
where:
\begin{eqnarray}
u_W^1 &=& \frac{1}{2}\psi_W -\frac{i}{2\pi} \int\frac{\bar
I_{W'}}{\bar I_{W}} \frac{\psi_{W'} }{(W'-W)}dW',  \label{35}
\\u_W^2 &=& \frac{1}{2}\psi_W +\frac{i}{2\pi} \int\frac{\bar
I_{W'}}{\bar I_{W}} \frac{\psi_{W'} }{(W'-W)}dW',  \label{36} \\
v_W^1 &=& \frac{1}{2}\phi_W -\frac{i}{2\pi} \int\frac{\bar
L_{W'}}{\bar L_{W}} \frac{\phi_{W'} }{(W'-W)}dW',  \label{37}
\\v_W^2 &=& \frac{1}{2}\phi_W +\frac{i}{2\pi} \int\frac{\bar
L_{W'}}{\bar L_{W}} \frac{\phi_{W'} }{(W'-W)}dW'. \label{38}
\end{eqnarray}
The most general stationary state $Z_W$ corresponding to the
energy $E_0+W$ can be written as a linear combination of the
states $Z_W^1$, $Z_W^2$ in Eqs (\ref{21}), (\ref{22}):
\begin{equation}  \label{39}
Z_{W}= \lambda Z_W^1+ \mu Z_W^2.
\end{equation}
By setting
\begin{equation}  \label{40}
Z_{W}= c\psi_0 + c_1u_W^1 + c_2u_W^2 + C_1v_W^1 + C_2v_W^2 ,
\end{equation}
we have:
\begin{eqnarray}
c &=& \frac{\lambda}{N'_W},  \label{41} \\
c_1 &=& \lambda\frac{\bar
I_{W}}{N'_W}(\frac{\varepsilon_{W}}{Q^2_W}-i\pi) +
\mu\frac{L_{W}}{Q_W},  \label{42} \\
c_2 &=& \lambda\frac{\bar
I_{W}}{N'_W}(\frac{\varepsilon_{W}}{Q^2_W}+i\pi) +
\mu\frac{L_{W}}{Q_W},  \label{43} \\
C_1 &=& \lambda\frac{\bar
I_{W}}{N'_W}(\frac{\varepsilon_{W}}{Q^2_W}-i\pi) -
\mu\frac{L_{W}}{Q_W},  \label{44} \\
C_2 &=& \lambda\frac{\bar
I_{W}}{N'_W}(\frac{\varepsilon_{W}}{Q^2_W}+i\pi) -
\mu\frac{L_{W}}{Q_W}. \label{45}
\end{eqnarray}
Majorana then focuses on particular stationary states with
$C_2=0$; from such condition he determines the coefficients
$\lambda, \mu$ and definitely obtains the expressions for $c, c_1,
c_2, C_1, C_2$. Quantities of particular interest are their
squared modulus:
\begin{eqnarray}
|c_1|^2 &=& \frac{\varepsilon_{W}^2+\pi^2(|I_W|^2-
|L_W|^2)^2}{\varepsilon_{W}^2+\pi^2Q^2},  \label{46} \\
|c_2|^2 &=&
\frac{4\pi^2|I_W|^2|L_W|^2}{\varepsilon_{W}^2+\pi^2Q^2},
\label{47}
\end{eqnarray}
while $|c_2|^2=1$ and $|c_1|^2 +|C_1|^2=1$. Now, let us consider
the usual approximation for which $I_W=I$, $L_W=L$ are constants.
In some applications the ratio:
\begin{equation}
p= p(\varepsilon)=\frac{|c_1|^2}{|c_2|^2}  \label{48}
\end{equation}
has a direct physical meaning, and is here given by
\begin{equation}  \label{49}
p(\varepsilon)=\frac{4\pi^2|I|^2|L|^2}{\varepsilon^2+\pi^2Q^4}.
\end{equation}
Its maximum value $p_0$ (obtained for $\varepsilon =0$) varies
between $0$ and $1$ (for $|I|^2=|L|^2$) and, putting $k =
|I|^2/|L|^2$, can be expressed as
\begin{equation}  \label{50}
p_0 =\frac{4k}{(k+1)^2}
\end{equation}
(note that $p_0(k)= p_0(1/k)$). It determines the integral:
\begin{equation}  \label{51}
\int p(\varepsilon) d \varepsilon =\pi^2 Q^2 p_0.
\end{equation}
In terms of the total disintegration probability of the unstable
state $\psi_0$,
\begin{equation}  \label{52}
\frac{1}{T}=\frac{2\pi}{\hbar}Q^2
\end{equation}
and the partial disintegration probabilities for transitions to
the states $\psi_W$ and $\phi_W$,
\begin{equation}  \label{53}
\frac{1}{T_1}=\frac{2\pi}{\hbar}|I|^2, \qquad \qquad
\frac{1}{T_2}=\frac{2\pi}{\hbar}|L|^2,
\end{equation}
with
\begin{equation}
\frac{1}{T}=\frac{1}{T_1}+ \frac{1}{T_2},
\end{equation}
\begin{equation}
\frac{1}{T_1}=\frac{k}{k+1}\,\,\,\frac{1}{T},\qquad
\frac{1}{T_2}=\frac{1}{k+1}\,\,\, \frac{1}{T},  \label{55}
\end{equation}
we have:
\begin{equation}
\int p(\varepsilon) d \varepsilon =
\frac{2\pi\hbar}{T}\frac{p_0}{4}=\frac{2\pi\hbar}{T}\frac{k}{(k+1)^2}=
\frac{2\pi\hbar}{T_1}\,\,\,\frac{1}{k+1}=\frac{2\pi\hbar}{T_2}\,\,\,\frac{k}{k+1}.
\label{56}
\end{equation}

\section{Application to $\alpha$-induced nuclear disintegration}

\noi The general theory outlined above was elaborated by Majorana
in order to describe the artificial disintegrations of nuclei by
means of $\alpha$-particles, with and without $\alpha$-absorption,
as recalled earlier. Majorana approaches the problem by
considering the simplest case with an unstable state (described by
$\psi_0$) of the system formed by a nucleus plus an
$\alpha$-particle, which spontaneously decays with the emission of
an $\alpha$-particle or a proton. For simplicity, he assumes that
such proton or $\alpha$-particle, coming from the disintegration
of $\psi_0$, is emitted as an $s$-wave and that the daughter
nucleus is always in its ground state. The initial system formed
by the parent nucleus plus the incoming $\alpha$-particle (in a
hyperbolic $s$-orbit) is described by the states $\psi_W$, while
the final one formed by the daughter nucleus and the free proton
(in an $s$-orbit) is described by the states $\psi_W$. The  state
$\psi_0$ is coupled to both $\psi_W$ and $\phi_W$ states by the
perturbation potential $H_p$ in Eqs. (\ref{14}), (\ref{15}).
Neglecting the motion of the nucleus, $\psi_W$ represents a
converging or diverging flux (number of particles per unit time)
of $\alpha$-particles equal to $1/2\pi\hbar$, while  $\phi_W$
represents an ingoing or outgoing flux of protons equal to
$1/2\pi\hbar$. Instead, the non-stationary states $u^1_W$ or
$u^2_W$ introduced above represent, at large distances, only
outgoing or ingoing fluxes of $\alpha$-particles (their intensity
being always $1/2\pi\hbar$ ), respectively, and similarly for
$v^1_W$ and $v^2_W$ describing outgoing or ingoing fluxes of
protons. The problem at face is that to study the scattering of (a
unitary flux per unit area of) $\alpha$-particles interacting with
the parent nucleus and to determine for what number of them the
nucleus disintegrates. To this end, Majorana considers a
stationary state, representing the incident plane wave plus a
diverging spherical wave of $\alpha$-particles plus a diverging
spherical wave of protons, as obtained by a sum of particular
solutions. Such solutions are not limited to those corresponding
to the parent nucleus plus $\alpha$-particles with non-vanishing
azimuthal quantum numbers, whose form is well known from the
theory of Coulomb scattering. They also account for incident
$\alpha$-particles with $\ell=0$ as well as a diverging wave of
$\alpha$-particles with $\ell=0$. Moreover, due to the coupling
between $\psi_0$ and $\phi_W$, the stationary state must also be
composed of an excited (at certain degree) $\psi_0$ state as well
as of a diverging wave of protons. Such a particular solution will
then have the form as in Eq. (\ref{40}) (with $C_2=0$) with the
coefficients taking the following values, except for a
proportionality factor:
\begin{eqnarray}
c&=&\frac{I_W}{N'_WQ_W},  \label{57} \\
c_1 &=&\frac{1}{N'_WQ_W}\left\{\varepsilon_W - i\pi\left(|I_W|^2 - |L_W|^2\right)\right\},  \label{58} \\
C_1 &=&-2\pi\frac{I_W\bar L_W}{N'_WQ_W},  \label{59} \\
c_2 &=&\frac{1}{N'_WQ_W}\left(\varepsilon_W + i\pi Q_W^2\right).
\label{60}
\end{eqnarray}
The coefficient $c_2$ can be determined from the condition that
the incoming flux of $\alpha$-particles is due to the incident
plane wave, this incoming flux being equal to $|c_2|^2/2\pi\hbar
.$ On the other hand, the number of $\alpha$-particles with
$\ell=0$ impinging on the nucleus per unit time is equal to the
flux through a circular cross section, normal to the propagation
direction of the wave, with radius $\lambda/2\pi$, $\lambda$ being
the wavelength of the $\alpha$-particles. Since the incident wave
represents a unit flux per unit area, we have:
\begin{equation}  \label{61}
\frac{|c_2|^2}{2\pi\hbar}=\pi\left(\frac{\lambda}{2\pi}\right)^2=
\left(\frac{\lambda^2}{4\pi}\right)= \frac{\pi\hbar^2}{M^2v^2},
\end{equation}
$M$ and $v$ being the mass and the velocity of the particles,
respectively. From this and from Eqs. (\ref{46}), (\ref{47}) the
value of $|c_2|^2$ and that of $|c_1|^2$ and $|C_1|^2$ may be
obtained:
\begin{eqnarray}
|c_1|^2 &=&\frac{\hbar\lambda^2}{2}\,\,\frac{\{\varepsilon^2
+\pi^2(|I|^2 -
|L|^2)^2\}}{\varepsilon^2 + \pi^2Q^4},  \label{62} \\
|C_1|^2 &=& \frac{\hbar\lambda^2}{2}\,\,\frac{4\pi^2 |I|^2
|L|^2}{\varepsilon^2 + \pi^2Q^4},  \label{63} \\
|c_2|^2 &=& \frac{\hbar\lambda^2}{2}.  \label{64}
\end{eqnarray}
The knowledge of the moduli of such coefficients suffices for the
problem at face, consisting in the study of only the frequency of
disintegration processes, disregarding scattering anomalies that
also depend on the phase of $c_1$. \\
The cross section $S(\varepsilon)$ for the disintegration process
is given by the outgoing proton flux, $|C_1|^2/2\pi\hbar$:
\begin{equation}  \label{65}
S(\varepsilon) =  \frac{\hbar\lambda^2}{4\pi}\,\,\frac{4\pi^2
|I|^2 |L|^2}{\varepsilon^2 + \pi^2Q^4}=
\frac{\hbar\lambda^2}{4\pi}p(\varepsilon)
\end{equation}
(by assuming, at first approximation, that $\lambda, I_W$ and
$L_W$ do not depend on $\varepsilon$). Thus $p(\varepsilon)$ is
interpretated as the probability that one particle with vanishing
azimuthal quantum number will induce one disintegration
($\lambda^2/4\pi$ is, in fact, the cross section for such
particles). This probability has a maximum for $\varepsilon =0$,
that is for the most favorable value of the energy, and $p_0$ can
reach the unit value for $k=1$ (see Eq. (\ref{50})). As pointed
out by Majorana, these means that if the state $\psi_0$ has the
same probability to emit a proton or an $\alpha$-particle, and the
energy of the incident  $\alpha$-particles takes its most
favorable value, then all the incident particles with vanishing
azimuthal quantum number will induce disintegration. \\
The analysis ends with the observation that when it is impossible
to directly measure the cross section $S(\varepsilon)$ for
particles with definite energy $E_0 + k + \varepsilon$, the
measurable quantity of interest is only $\int S(\varepsilon)
d\varepsilon$ that is given, according to Majorana, by the
expression
\begin{equation}  \label{66}
\int S(\varepsilon) d\varepsilon =  \frac
{\lambda^2}{4\pi}\,\,\frac{\pi\hbar}{2T}\,p_0=
\frac{\hbar\lambda^2}{2T}\,\,\frac{k}{(k+1)^2}.
\end{equation}
A complete theory for the artificial disintegration of nuclei
induced by $\alpha$-particles, which is fully accessible to
experimental investigation, is thus finally provided.

\section{Conclusions}

\noi In the historical reconstruction of the path towards the
understanding of the structure of atomic nuclei, it is usually
paid little attention to the experiments performed on the
artificial disintegration of nuclei by bombardment with
$\alpha$-particles. This is probably due to the (presumed) lack of
a comprehensive theoretical interpretation of them and its
consistent inclusion in the generally accepted framework of
Quantum Mechanics applied to Nuclear Physics. In this respect, the
original results from Chadwick and Rutherford in 1929, on the
peculiarities of proton emission in the artificial disintegration
of some nuclei, were puzzling when assuming the capture of the
incident $\alpha$-particle by the atomic nucleus. Only later, in
1930-31, with the fundamental contribution of Gamow, it was
recognized that such disintegrations could take place even without
the capture of the $\alpha$-particles, with the assumption that
the protons and $\alpha$-particles contained in a nucleus are in
definite energy levels. Thus, in general, the energy spectrum of
the ejected protons resulted to be the superposition of a
continuous spectrum and a line, as observed experimentally. Once
realized this, the next step should have been the elaboration of a
complete theory of such processes in the general framework of
Quantum Mechanics. However, as a matter of fact, this was not even
considered, as far as we know, by the main theorists who were
working on that. Only recently we have realized that such an issue
was effectively studied by Majorana in his personal researches,
although he didn't publish anything on it. As shown in the
previous pages, this was simply achieved by following the dynamics
of a state resulting from the superposition of a discrete state
with a continuous one, interacting between them through a term
like in Eq. (\ref{1}). The physical problem at hand was, then,
approached by considering the simplest case with an unstable state
of the system formed by a nucleus plus an $\alpha$-particles,
which spontaneously decays with the emission of an
$\alpha$-particle and a proton. As a result, Majorana succeeded in
obtaining, among the other predictions, the explicit expression
for the integrated cross section of the nuclear process, which was
the direct measurable quantity of interest in the experiments. It
should be noted that quasi-stationary states (the superposition of
discrete and continuous states) were earlier introduced by Rice in
a completely different context, that is the phenomenon of
predissociation in molecules \cite{[Rice]}. We do not know if
Majorana was aware of the Rice's papers, but we have to observe
that, disregarding the dissimilar theoretical elaborations, even
the notations used by the two authors appear very different
between them. Instead it is quite remarkable that another name is
usually associated to the study of the quantum interference
phenomenon between a discrete level and a continuum, i.e. the name
of U. Fano. Indeed, the fundamental paper cited by almost everyone
in this respect was that written by Fano at the beginning of 1935
\cite{[Fano]}, where he investigated the stationary states with
configuration mixing under conditions of autoionization, when
interpreting the strange looking shapes of spectral absorption
lines of atoms in the continuum. The important result achieved by
this author in an Atomic Physics framework was later related with
the resonant scattering of a slow neutron in a nucleus, that is
the so-called ``shape resonances" found by Fermi
\cite{[Bianconi]}. The starting point of Fano, namely the writing
of the perturbed eigenfunction of the system considered, as in the
last equation on page 156 of Ref. \cite{[Fano]}, is surprisingly
similar to what introduced by Majorana in his work of 1930. It is
even remarkable to note the curious coincidence that Fano
elaborated his theory just after few time when he moved in 1934-35
to Rome to work with Fermi \cite{[Bianconi]}. Then it cannot be
excluded a certain influence of Majorana, even indirectly (in his
paper, Fano acknowledged the help given to him by Fermi), on this
issue. In any case, it is certainly astonishing that a complete
theory, with application of the concept of the quasi-stationary
states to a particular Nuclear Physics issue, was achieved by
Majorana as early as in 1930, well before that the Fermi group
decided to switch from Atomic Physics matters to those related to
the structure of nuclei. Some other surprises we should expect
from further studies of the unpublished papers of the Italian
scientist.


\end{document}